\documentclass[runningheads]{llncs}

\usepackage[T1]{fontenc}

\usepackage{ifthen}
\newboolean{FinalVersion}
\setboolean{FinalVersion}{true}

\usepackage{amsmath,amssymb,amsfonts}
\usepackage{algorithmic}
\usepackage{graphicx}
\usepackage{xcolor}
\usepackage{acro}
\usepackage{xspace}
\usepackage{multirow}
\usepackage{caption}
\usepackage{subcaption}
\usepackage{wrapfig}
\usepackage{graphicx}
\usepackage{listings}
\usepackage{bbding}

\usepackage{tabularx}
\newcolumntype{Y}{>{\raggedright\arraybackslash}X}
\usepackage{array}       
\usepackage{multirow}    
\usepackage{booktabs}    

\usepackage[ruled,vlined]{algorithm2e}

\usepackage{color}
 
\providecommand{\shortcite}[1]{\cite{#1}}

\begin{document}

\title{Disaggregated Design for GPU-Based Volumetric Data Structures}

\author{Massimiliano Meneghin\inst{1}\orcidID{0000-0002-1295-9062} \and
\\ Ahmed H. Mahmoud\Envelope\inst{2}\orcidID{0000-0003-1857-913X}}

\authorrunning{Meneghin and Mahmoud}
\titlerunning{Disaggregated Design for GPU-Based Volumetric Data Structures}

\institute{Autodesk Research, Milan, MI 20124, Italy\\
\email{massimiliano.meneghin@autodesk.com}\\
\and
MIT CSAIL, Cambridge, MA 02139, USA\\
\email{ahdhn@mit.edu}}

\maketitle

\begin{abstract}

Volumetric data structures typically prioritize data locality, focusing on efficient memory access patterns. This singular focus can neglect other critical performance factors, such as occupancy, communication, and kernel fusion. We introduce a novel \emph{disaggregated} design that rebalances trade-offs between locality and these objectives---reducing communication overhead on distributed memory architectures, mitigating register pressure in complex boundary conditions, and enabling kernel fusion. We provide a thorough analysis of its benefits on a single-node multi-GPU Lattice Boltzmann Method (LBM) solver. Our evaluation spans dense, block-sparse, and multi-resolution discretizations, demonstrating our design's flexibility and efficiency. Leveraging this approach, we achieve up to a $3\times$ speedup over state-of-the-art solutions.

\keywords{Data layout \and Parallel \and Simulation \and GPU \and LBM.}
\end{abstract}

\section{Introduction}
\label{sec:intro}

Since the 2000s, the \emph{memory wall}~\cite{Wulf:HTM:1995,Asanovic:VPC:2009} has underscored the critical importance of data locality optimizations in computational tasks. This challenge is especially acute in memory-bound volumetric physics simulations, prompting research into strategies such as blocking~\cite{Endo:2018:ART}, time-tiling~\cite{Wonnacott:2000:UTS}, polyhedral optimizations~\cite{Bondhugula:2008:APA}, and cache-oblivious methods~\cite{Frigo:2012:COA}.

Although GPUs provide high memory bandwidth, achieving peak performance also requires addressing occupancy, load balancing, synchronization overhead, and data movement. Traditionally, volumetric data structures are designed to improve data locality first, with other optimizations (e.g., overlapping computation and communication~\cite{Meneghin:2022:NAM}, time skewing~\cite{Wonnacott:2000:UTS}, kernel fusion~\cite{Wang:2020:AMP}, tiling~\cite{Tran:2017:POO}) introduced afterward. Because these methods are not considered during data structure design, significant performance opportunities are lost. We argue that additional objectives should be incorporated into volumetric data structure design. While data locality remains essential, selectively compromising it can improve end-to-end performance by addressing other goals.

In this paper, we propose \emph{disaggregated design} for volumetric data structures, which balances multiple performance objectives by:

\begin{enumerate} \item \textbf{Grouping voxels based on desired properties:} Rather than relying solely on spatial locality, we cluster voxels according to the traits most relevant for performance. \item \textbf{Applying traditional data locality optimizations within each group:} Within these groups, we still exploit locality as appropriate while addressing other performance targets. \end{enumerate}

We evaluate our \emph{disaggregated} approach on a LBM fluid solver running on single- and multi-GPU systems, achieving:

\begin{itemize} 
    \item A zero-copy multi-GPU implementation that overlaps computation and communication for dense discretizations, \textbf{minimizing transfer overhead} and delivering up to a $3\times$ speedup over state-of-the-art solutions. 
    \item A disaggregated interface and layout for block-sparse data structures that \textbf{reduce high register pressure} in complex boundary conditions (e.g., regularized LBM~\cite{Latt:2008:regularized}), achieving up to a $2\times$ speedup over naive implementations without extra boundary-data storage. 
    \item A multi-resolution grid representation that \textbf{maximizes kernel fusion} in regions unaffected by neighboring cells of different sizes, yielding up to a 26\% performance improvement on a single GPU. 
\end{itemize}

In Section~\ref{sec:theConcept}, we formalize the disaggregated design methodology. Sections~\ref{sec:dis_sten_pattern}, \ref{sec:dis_map_pattern}, and \ref{sec:dis_mres_pattern} apply this approach to dense, sparse, and multi-resolution volumetric grids, respectively. Section~\ref{sec:eval} presents our evaluation using LBM solvers across multiple GPU architectures. We discuss related work in Section~\ref{sec:background} and conclude in Section~\ref{sec:conclusion}.
\section{Disaggregated Design Method}
\label{sec:theConcept}

Voxel-based representations, derived from Cartesian discretization, include \textbf{dense} (every voxel in a multidimensional interval is allocated), \textbf{sparse} (an irregular subset of the interval), and \textbf{multi-resolution} (voxels of different sizes in a single interval). Traditional design efforts have focused on data locality to reduce the growing gap between compute speed and memory latency. However, other performance-critical optimizations---e.g., minimizing communication overhead, reducing register pressure, and maximizing kernel fusion---are equally important. To address these, we introduce \emph{disaggregated design}, a multi-objective method for volumetric data structures defined as follows:

\begin{definition} Given an optimization objective $\Phi$ to be considered alongside data locality, a disaggregated design maps data over a voxelized domain into a 1D memory space in four steps: \begin{enumerate} \item \textbf{Definition:} Identify properties $\mathcal{P}{1}, \ldots, \mathcal{P}{n}$ influencing $\Phi$. \item \textbf{Classification:} Group voxels $\mathcal{G}_1, \ldots, \mathcal{G}_n$ based on those properties. \item \textbf{Mapping:} Within each group $\mathcal{G}_i$, map voxel data to memory using classical data-locality optimizations. \item \textbf{Operations:} Apply group-specific operations to maximize $\Phi$. \end{enumerate} \end{definition}

By incorporating objectives beyond locality (\textbf{definition}), grouping voxels accordingly (\textbf{classification}), and then applying classical optimizations locally (\textbf{mapping}), we enable targeted \textbf{operations} that yield higher end-to-end performance. Success depends on whether gains from optimizing $\Phi_i$ outweigh potential drawbacks, e.g., sub-optimal locality since locality may suffer if inter-group optimizations are underutilized or increased complexity since additional indexing is needed for separate groups.

We study disaggregated design within a \textbf{for-each} data-parallel pattern applying a side-effect-free function to each voxel. Under this model, we consider three compute patterns: (1)~\emph{Map Pattern} where each voxel depends only on local data, (2)~\emph{Uniform Stencil Pattern} where each voxel queries its neighbors (e.g., convolution), and (3)~\emph{Multi-resolution Stencil Pattern} where varying voxel sizes require neighbor access at different resolutions. 

In the following, we apply disaggregated design to dense (Section~\ref{sec:dis_sten_pattern}), sparse (Section~\ref{sec:dis_map_pattern}), and multi-resolution representations (Section~\ref{sec:dis_mres_pattern}). We then detail its performance impact on a fluid simulation solver in Section~\ref{sec:eval}. 
\section{Disaggregation on a Dense Domain}
\label{sec:dis_sten_pattern}

\begin{figure*}[t]
	\centering
	\subfloat[\label{fig:jacobiParition}]{
		\includegraphics[height=.17\linewidth]{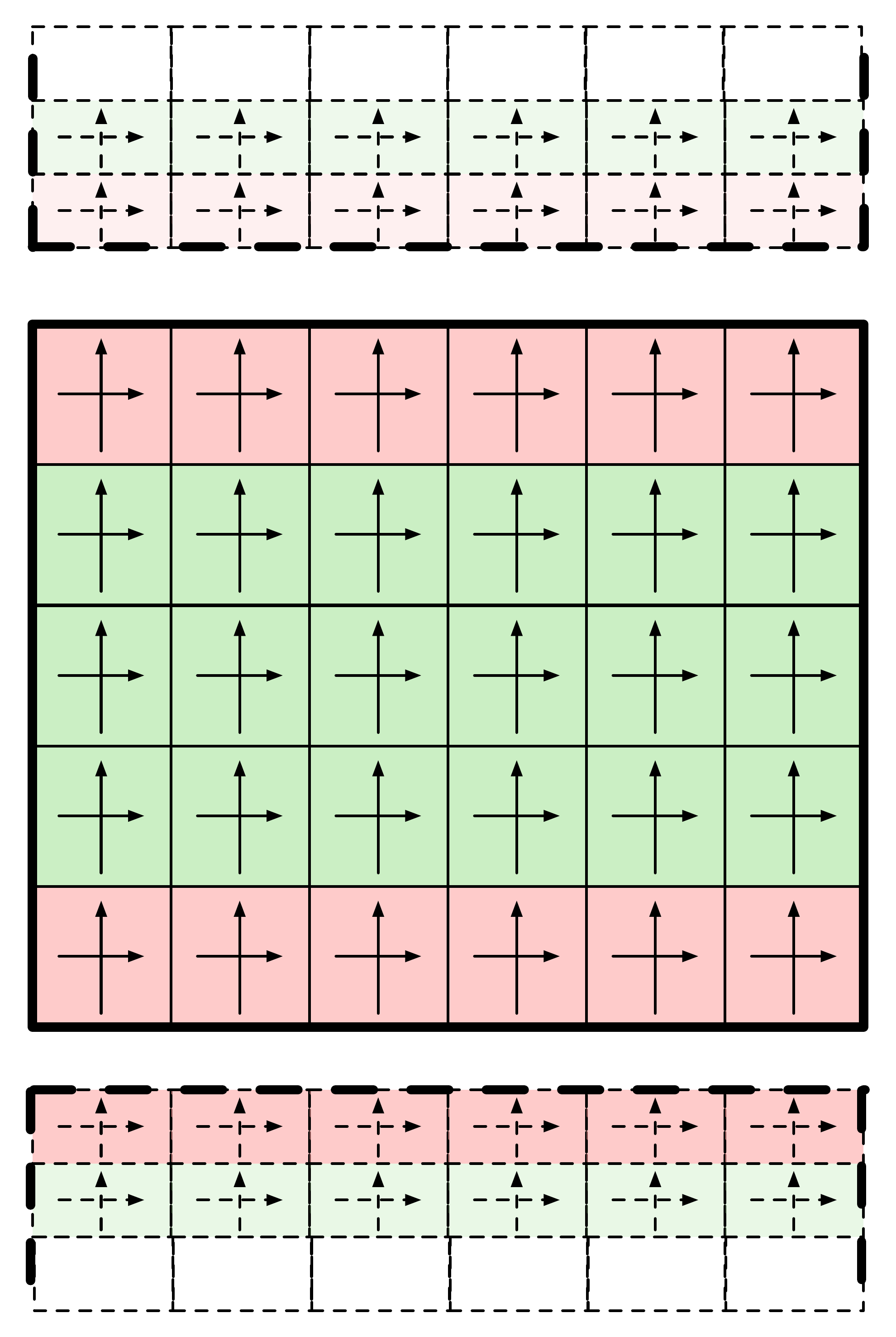}
	}\hfill
	\subfloat[AoS \label{fig:jacobiAoS}] {
		\includegraphics[height=.17\linewidth]{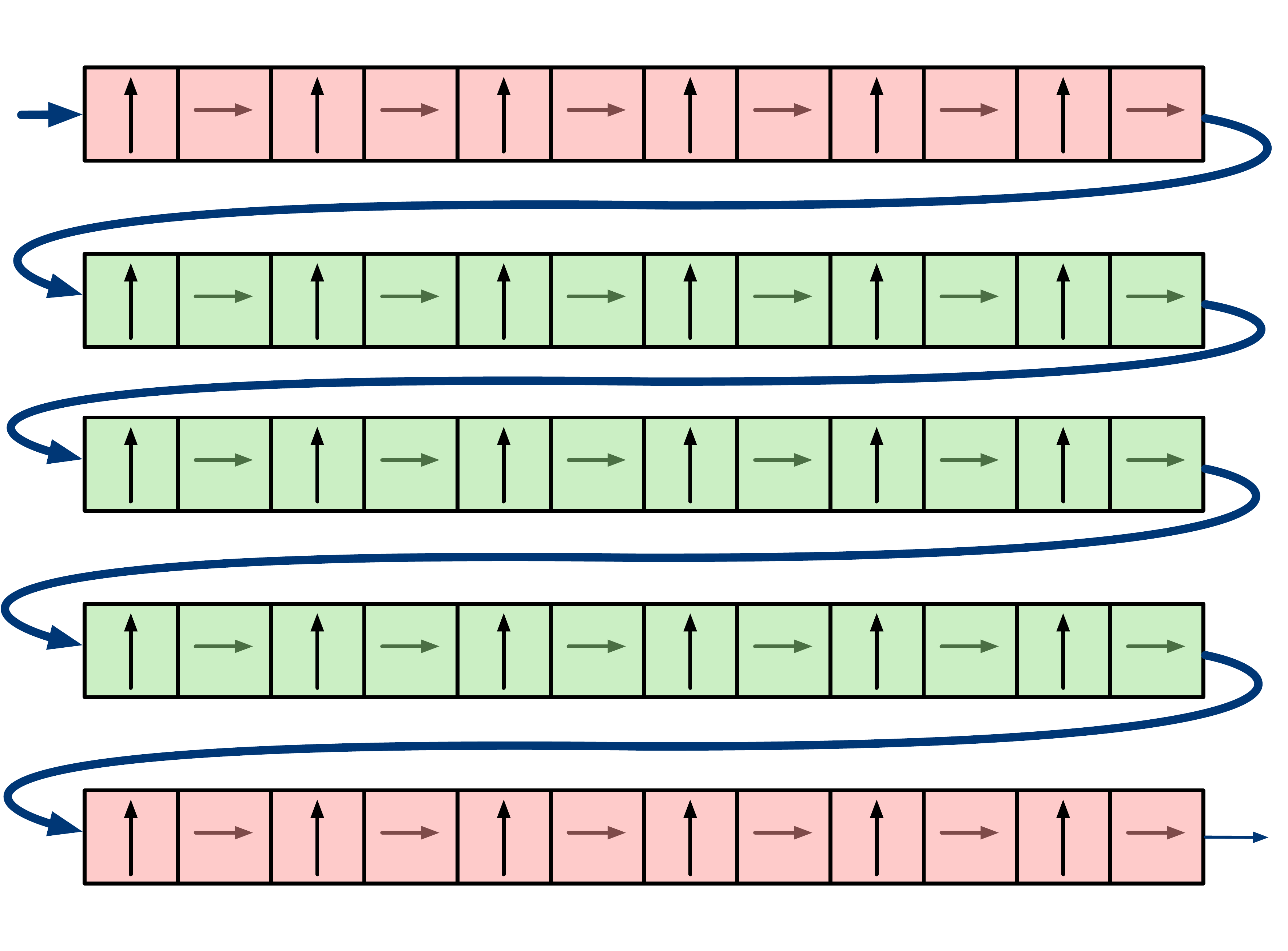}
	}\hfill
	\subfloat[SoA \label{fig:jacobiSoA}] {
		\includegraphics[height=.17\linewidth]{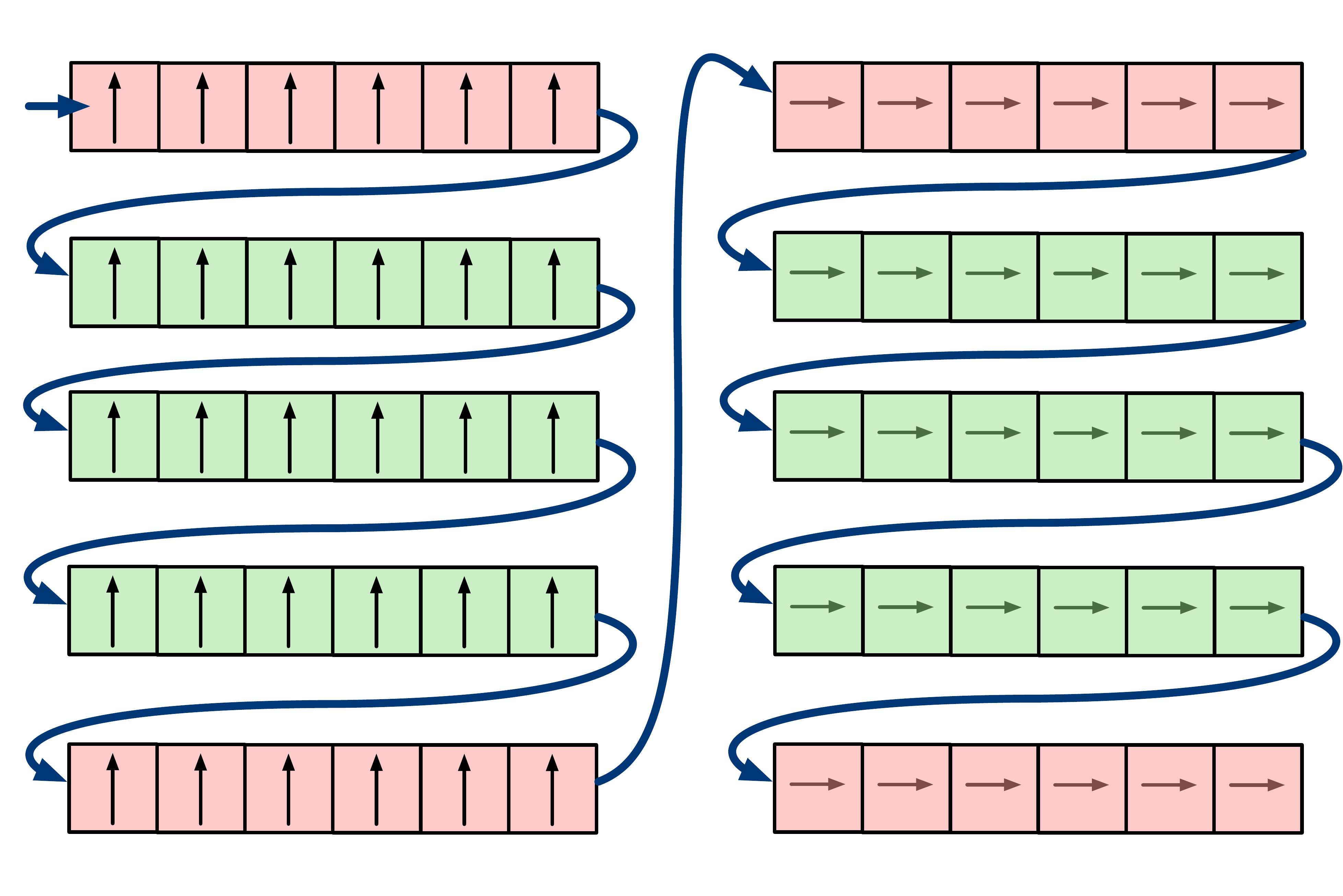}
	}\hfill
	\subfloat[Disaggregated \label{fig:jacobiDisg}] {
		\includegraphics[height=.17\linewidth]{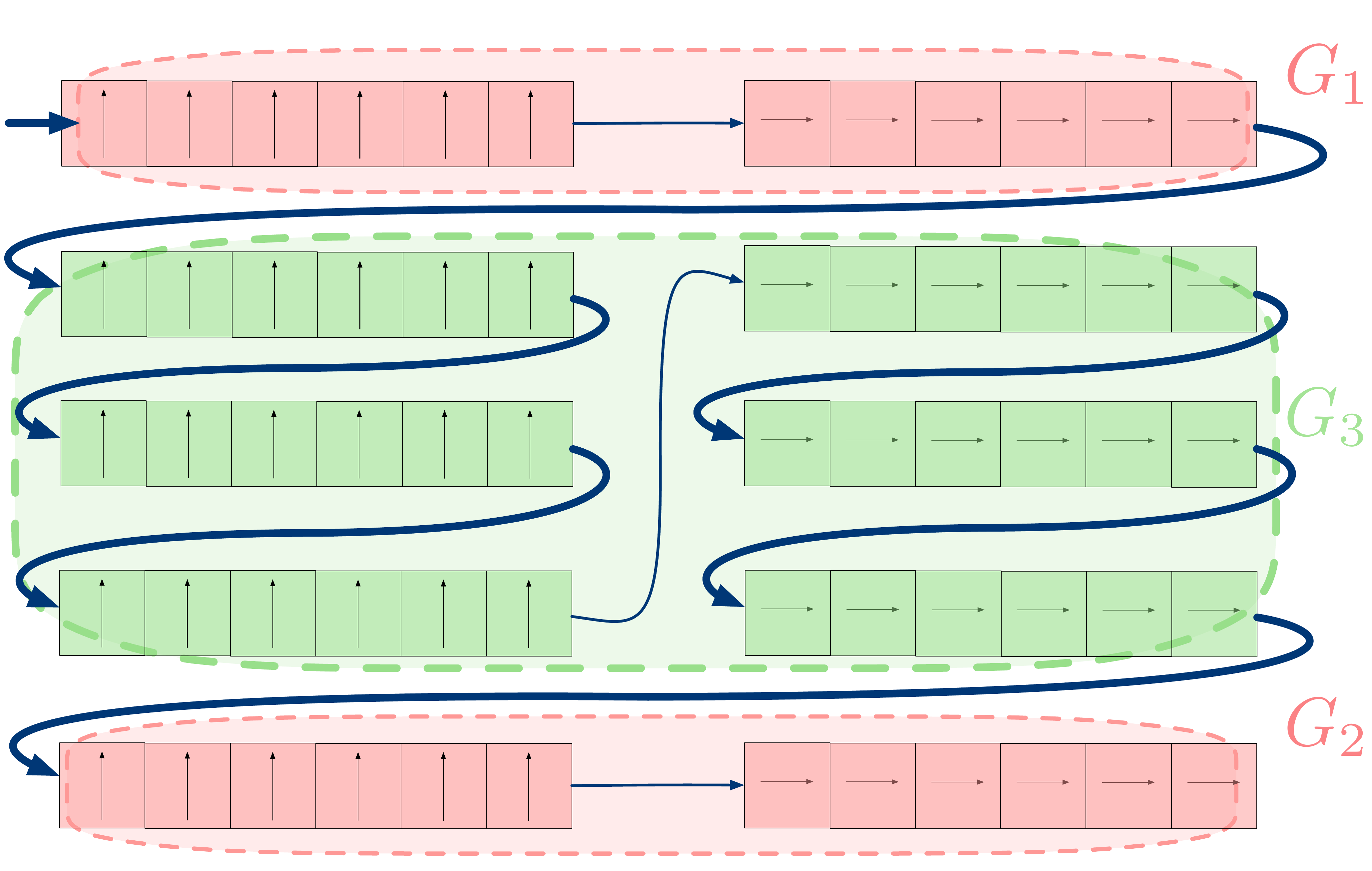}
	}
	\caption{\label{fig:jacobi}
		Illustration of a five-point stencil on a two-component vector field in a 2D domain with partitioning along one axis (a) and three different memory layouts (b,c,d).
	}
\end{figure*}

In this setup, a dense grid is divided across multiple GPUs, each handling one partition. In stencil computations that require neighbor data, fetching data directly from neighboring GPUs each iteration is highly inefficient due to communication overhead. To mitigate this, each partition maintains a \emph{halo region} (Figure~\ref{fig:jacobiParition}). Synchronizing these halos (the \emph{halo update}~\cite{Meneghin:2022:NAM}) can significantly add to execution time if performed before every stencil step. \emph{Overlapping Computation and Communication} (OCC) addresses this by dividing the stencil update into two phases. First, \textbf{private} voxels (relying only on local data) are processed while the halo is updated in parallel. Then, \textbf{shared} voxels (requiring neighbor data) are computed. This hides communication costs and improves scalability on multi-GPU systems.

We adopt a communication model~\cite{Culler:1993:LTA} with a constant setup time, $t_{setup}$, plus a term proportional to message size ($size(\text{msg})$) and the interconnect throughput, $b_{com}$, such that
$
	t_{send}(\text{msg}) = t_{setup} + \frac{size(\text{msg})}{b_{com}}.
$
If exchanged data resides in disjoint memory regions, multiple transfers are required.

We consider a 2D stencil on a vector field, where each point stores a 2D vector. Two common layouts are \emph{Array-of-Structures} (AoS) and \emph{Structure-of-Arrays} (SoA). SoA typically yields better coalesced GPU memory access~\cite{Wittmann:2013:COD}. For a grid of size $d_x \times d_y$, partitioned along one dimension (Figure~\ref{fig:jacobiParition}), the halo-update time for a generic partition can be approximated as:

\begin{equation}
	\label{eq:halo_update_model_1D}
	t_{halo\_update} = \alpha t_{setup} + \beta \frac{size(T)}{b_{com}}
\end{equation}

where $\alpha$ is the number of transfer operations, and $\beta$ the total number of elements sent. With a 1D decomposition, $\alpha \ge 2$ (upper and lower neighbors), and $\beta = 2 \cdot d_y$ for shared boundary elements.

\begin{wraptable}[12]{r}{0.4\textwidth}
	\vspace{-1em}
	\centering
	\small
	\begin{tabular}{r|ccc}
		                   & $\boldsymbol{\alpha}$ & $\boldsymbol{\beta}$ & \textbf{Coalesced} \\
		\hline
		\textbf{AoS}       & 2                     & $2 \cdot d_x$        & No                 \\
		\textbf{SoA}       & 4                     & $2 \cdot d_x$        & Yes                \\
		\textbf{Disag SoA} & 2                     & $2 \cdot d_x$        & Yes                \\
	\end{tabular}
	\caption{\label{table:dissComparisonJacobi}
		Comparison of disaggregated, AoS, and SoA layouts for a five-point stencil using the model in Eq.~\ref{eq:halo_update_model_1D}. The 2D domain has dimensions $d_x \times d_y$, with a 1D partition along the y-axis.}
\end{wraptable}

Using AoS (Figure~\ref{fig:jacobiAoS}) keeps shared-voxel data contiguous, minimizing $\alpha$ to 2, but it breaks coalesced GPU access~\cite{Wittmann:2013:COD}. Conversely, SoA (Figure~\ref{fig:jacobiSoA}) preserves coalesced accesses but splits data, increasing $\alpha$ to 4. This increase occurs because the 2D components are stored non-contiguously in memory, as illustrated by the four distinct regions in Figure~\ref{fig:jacobiSoA}. To reduce the number of communication operations, the data would need to be copied into a contiguous buffer.

To reduce $\alpha$ while retaining coalesced accesses, we apply \emph{disaggregated design}. We define $\mathcal{P}_1$ and $\mathcal{P}_2$ to enforce contiguous mapping for voxels shared with the upper ($\mathcal{G}_1$) and lower ($\mathcal{G}_2$) neighbors, while remaining private voxels form $\mathcal{G}_3$ (Figure~\ref{fig:jacobiDisg}). We then map each group in SoA format, preserving $\mathcal{P}_1$ and $\mathcal{P}_2$.

Table~\ref{table:dissComparisonJacobi} compares these layouts, showing that the \emph{disaggregated SoA} merges the benefits of AoS (minimal transfers) with SoA’s coalesced memory access.
\section{Disaggregation on a Sparse Domain}
\label{sec:dis_map_pattern}

When the region of interest in a simulation domain is significantly smaller than the full domain, dense representations become inefficient. In these scenarios, \emph{sparse} representations are preferred, allocating data only for actively used voxels and thus conserving memory and compute resources. A common use case in sparse domains involves handling boundary conditions in physics solvers, where computations on each voxel may vary based on its boundary type. For instance, in computational fluid dynamics, no-slip conditions are enforced on boundary voxels at walls, whereas interior (non-boundary) voxels typically follow the Navier--Stokes equations.

\begin{table}[t]
    \centering
    \small
    \begin{tabular}{c|ccccc}
                                         & \textbf{\# Kernels}       & \textbf{\# Blocks} & \textbf{\# Registers} & \textbf{Storage}                                      & \textbf{Indexing }      \\
        \midrule \midrule
        Naive                            & 1                         & $n_{b}+n_{nb}$     & $r_{b}$               & $s_w  n_{nb} b_{\text{size}}$                         & Direct                  \\
        \midrule
        \multirow{2}{*}{Disag - Bitmask} & \multirow{2}{*}{2}        & $n_{b}+n_{nb}$     & $r_{b}$               & \multirow{2}{*}{$s_i (n_{b}+n_{nb}) b_{\text{size}}$}
                                         & \multirow{2}{*}{Indirect}                                                                                                                                \\
                                         &                           & $n_{b}+n_{nb}$     & $r_{nb}$              &                                                       &                         \\
        \midrule
        \multirow{2}{*}{Disag - Mem}     & \multirow{2}{*}{2}        & $n_{b}$            & $r_{b}$               & \multirow{2}{*}{$0$}                                  & \multirow{2}{*}{Direct} \\
                                         &                           & $n_{nb}$           & $r_{nb}$              &                                                       &                         \\
    \end{tabular}
    \caption{\label{table:dissBlock}
        Comparison of the disaggregated design vs.\ a naive approach for a map pattern involving complex boundary conditions in a block-sparse representation. \# Kernels is the number of kernels launched; \# Blocks is the number of blocks per kernel; \# Registers is the register usage; Storage quantifies additional space needed for boundary metadata ($s_w$ is the memory size per boundary voxel, $b_{\text{size}}$ is the number of voxels in a block, and $s_i$ is the size of the indexing type).}
\end{table}

The ratio of boundary to total voxels is often small, as boundaries generally scale with surface area rather than volume. Here, we assume a block-sparse representation (commonly managed by space-filling curves) though other layouts are possible. The computational load on boundary voxels varies based on their type, introducing a few challenges for efficient GPU implementations:

\begin{itemize}
    \item \textbf{Register pressure:} Additional registers may be needed for boundary computations, decreasing kernel occupancy.
    \item \textbf{Memory overhead:} Managing boundary conditions often requires per-voxel metadata, increasing memory requirements.
\end{itemize}

Let $r_{\text{nb}}$ be the resource needs for \underline{n}on-\underline{b}oundary computations and $r_{\text{b}}$ for \underline{b}oundary computations. We focus on the practical case $r_{\text{b}} > r_{\text{nb}}$, which can degrade performance by reducing occupancy and increasing memory usage.

\vspace{-1em}
\subsubsection{Naive Approach}
A naive solution launches a single kernel for all voxels. Because boundary logic is included, the kernel’s resource demand is
$
    r = \max(r_{\text{nb}}, r_{\text{b}}) = r_{\text{b}}.
$
Even though most voxels only need $r_{\text{nb}}$, the kernel is constrained by $r_{\text{b}}$. This often leads to suboptimal occupancy and high memory usage, i.e., allocating a full buffer for all voxels, even though only some are boundary voxels.

\vspace{-1em}
\subsubsection{Disaggregated Approach}
Sparse domains typically exhibit heterogeneous workloads, with boundary voxels requiring significantly more resources. The \emph{disaggregated design} alleviates this by separating the domain into two groups:

\begin{itemize}
    \item \textbf{Boundary Group} ($\mathcal{G}_{\text{b}}$): Blocks containing at least one boundary voxel.
    \item \textbf{Non-Boundary Group} ($\mathcal{G}_{\text{nb}}$): Blocks with only non-boundary voxels.
\end{itemize}

This separation allows two specialized kernels, each optimized for its target group. Next, we consider two implementations of this concept:

\vspace{-1em}
\paragraph{Memory-Based Grouping:}
All boundary blocks are contiguous in memory, followed by non-boundary blocks. Each group is processed by a separate kernel, removing the need for runtime checks on block type and simplifying memory access.

\vspace{-1em}
\paragraph{Bitmask-Based Grouping:}
In this implementation, we use a bitmask at runtime to distinguish block types. Since the spans of the two groups are no longer contiguous, both kernels must execute over the entire domain. Memory for boundary-specific data is allocated using \emph{indirect indexing}, where a unique identifier is assigned to each voxel. This identifier maps boundary voxels to their metadata, which is stored in a contiguous buffer.

Table~\ref{table:dissBlock} summarizes these approaches. We examine the performance results and trade-offs of these two implementations in Section~\ref{sec:registerPressure}.

\section{Disaggregation on a Multi-resolution Domain}
\label{sec:dis_mres_pattern}

\begin{figure*}[t]
    \centering
    \subfloat[Multi-res domain\label{fig:mres:domain}] {
        \includegraphics[height=.12\linewidth]{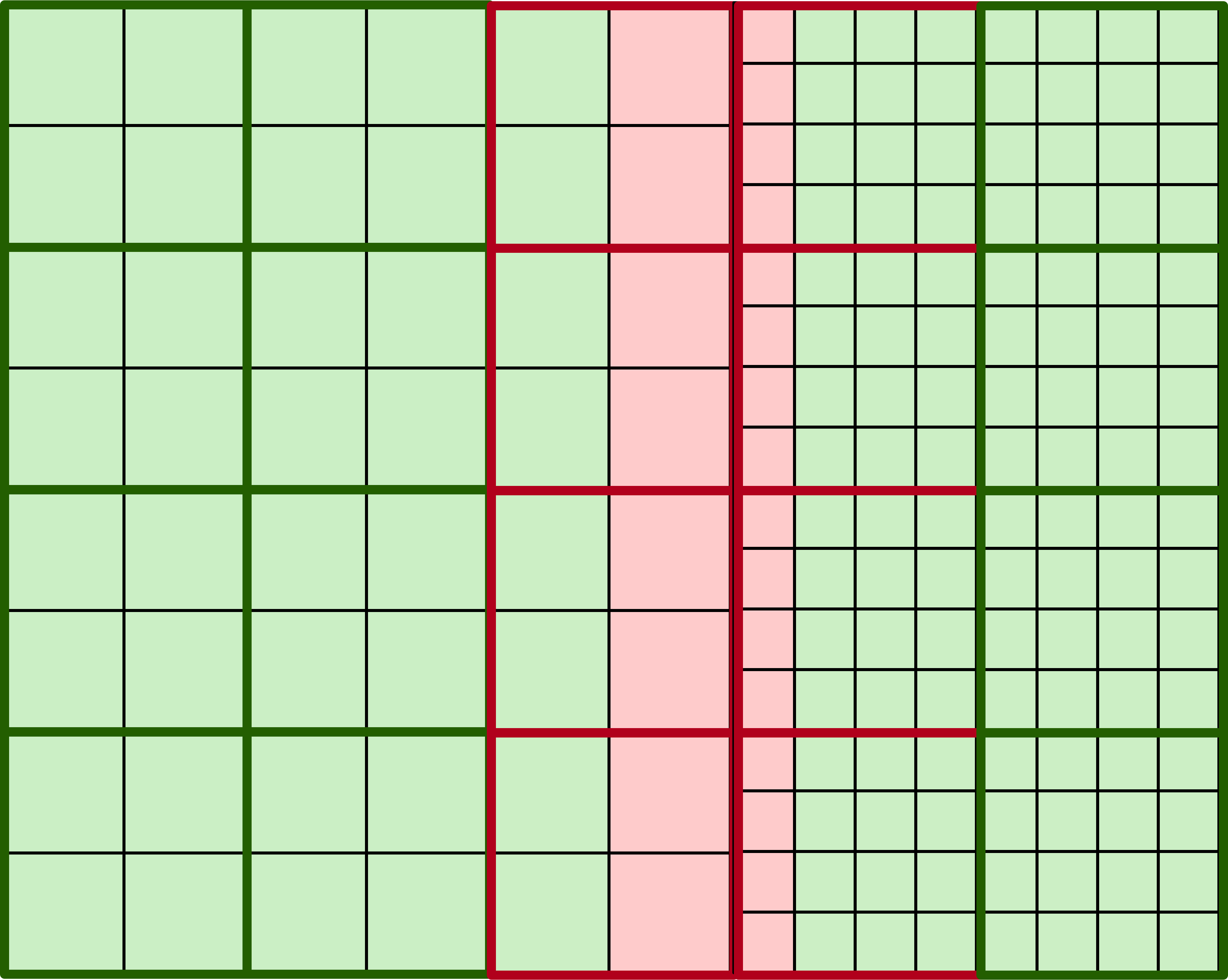}
    }\hfil
    \subfloat[Comp.\ Graph\label{fig:mres:naive}]{
        \includegraphics[height=.12\linewidth]{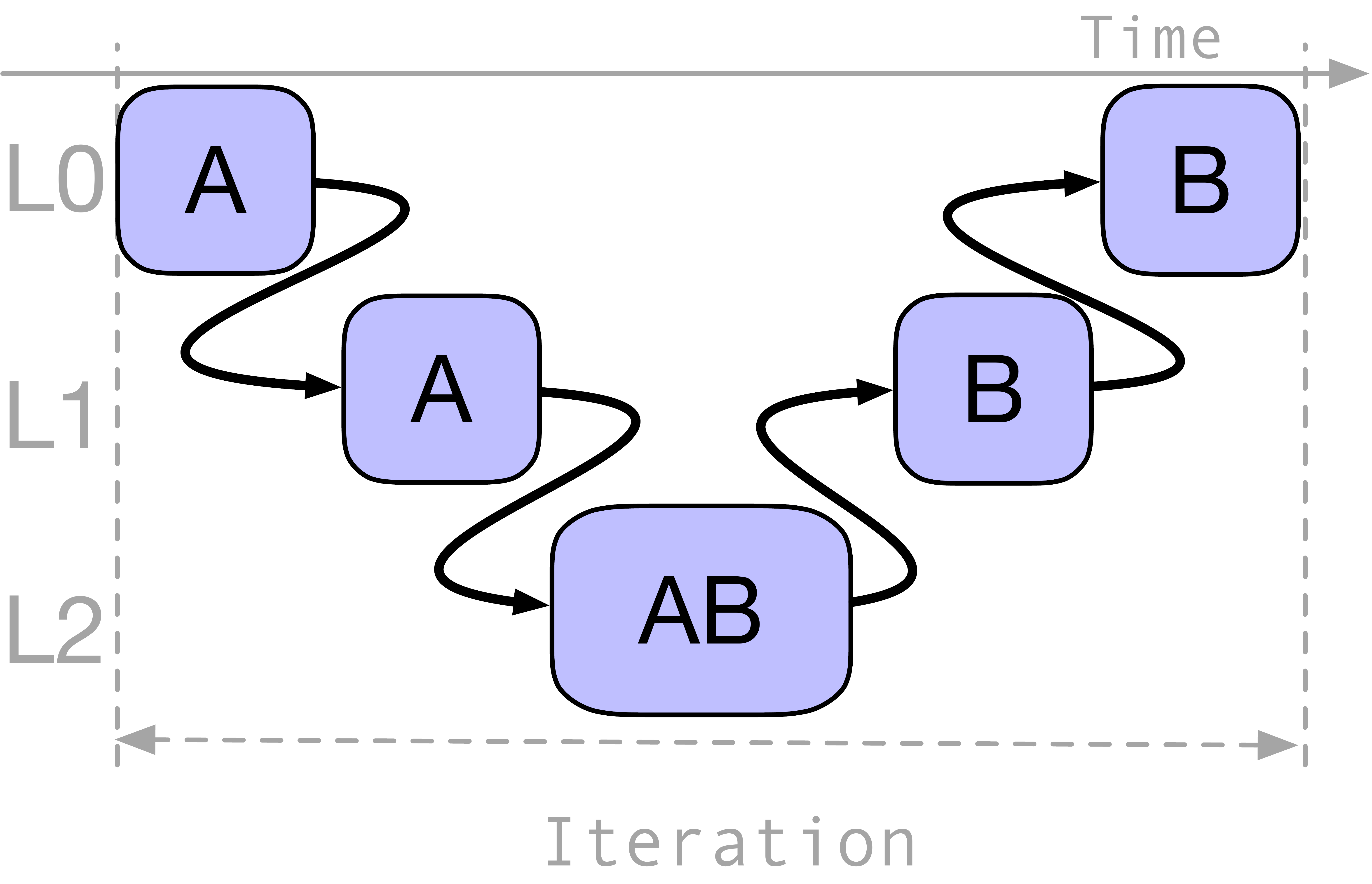}
    }\hfil
    \subfloat[Disag.\ Comp.\ Graph\label{fig:mres:disg}]{
        \includegraphics[height=.12\linewidth]{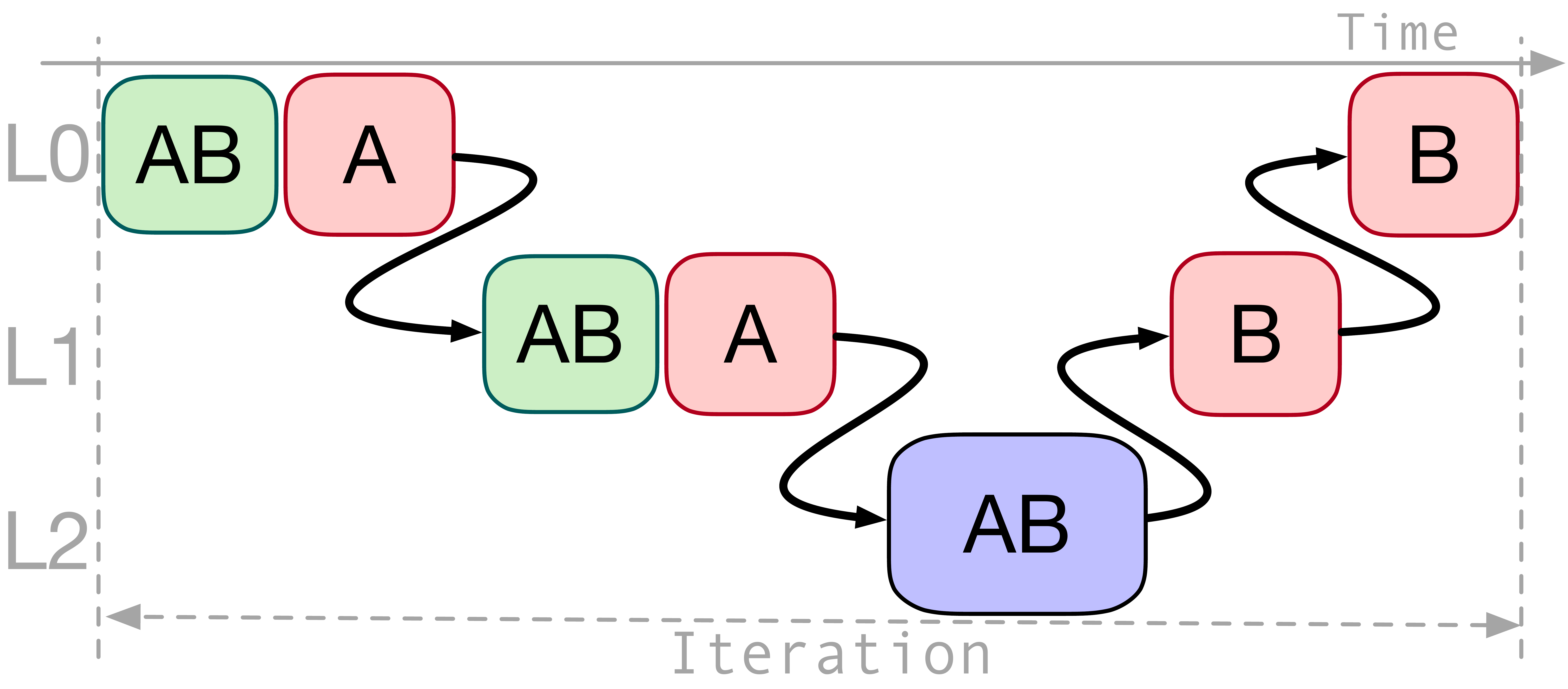}
    }
    \caption{\label{fig:mres}
    A multi-resolution domain with three levels (a) where red blocks lie near resolution transitions, and green blocks are farther away. The computational graph (b) shows dependencies between two kernels: both run on red and green blocks, but kernel B waits for cross-level boundary data. Fusing kernels is traditionally feasible only at the finest resolution level. In the disaggregated approach (c), green blocks fuse computations at any level, while red blocks execute two kernels sequentially once boundary data is available, reducing iteration time.
    }
\end{figure*}

Multi-resolution data structures handle voxels of varying sizes in one domain (Figure~\ref{fig:mres}), supporting both \emph{intra-level} stencil operations (within a single resolution) and \emph{cross-level} interactions (between adjacent resolutions).

During each time step in multi-resolution solvers~\cite{Lagrava:2012:AIM}, only voxels near resolution boundaries require cross-level communication. Figure~\ref{fig:mres} distinguishes green (intra-level only) from red (cross-level) voxels. Due to producer/consumer dependencies, iterations are typically split into two steps and can be fused only at the finest level (Figure~\ref{fig:mres:naive}).

Using the disaggregated design, we improve memory throughput by maximizing kernel fusion for intra-level computations. Voxels far from resolution jumps do not need cross-level data, so their iterations can fuse at any resolution level. To formalize this, we define a discrete distance property, $\mathcal{P}_{d}$, measuring how close a voxel is to a resolution jump (distance 0 indicates immediate proximity). Each resolution level is partitioned into: (1) $\mathcal{G}_{\text{i}}$: Blocks where all voxels have distance $\ge 1$, allowing fully fused operations, and (2) $\mathcal{G}_{\text{c}}$: Blocks with at least one voxel at distance 0, requiring separate cross-level and intra-level steps.

We apply a standard memory locality layout to each group within each level. Under disaggregation, $\mathcal{G}_{\text{i}}$ blocks use fused kernels across resolution levels, whereas $\mathcal{G}_{\text{c}}$ blocks wait for boundary data (Figure~\ref{fig:mres:disg}). This approach minimizes memory pressure for $\mathcal{G}_{\text{i}}$ while preserving accurate cross-level operations for $\mathcal{G}_{\text{c}}$.

\section{Evaluation and Discussions}
\label{sec:eval}
We evaluate the disaggregated design method using a fluid dynamics simulation based on LBM on dense, sparse, and multi-resolution grids. We selected LBM as a representative application since it could benefit from many of the objectives our design method targets. LBM models the time evolution of \emph{velocity distribution functions} ($f_i$) along discrete lattice directions $e_i = (e_1, \ldots, e_q)$. In 3D, we use lattices with 19 (D3Q19) or 27 (D3Q27) directions. Each $f_i$ value, or \emph{population}, evolves through a \emph{collide-and-stream} process. \emph{Collision} is a nonlinear, local operation that modifies $f_i$ at each lattice point. Here, we employ the BGK single-relaxation-time model for the collision~\cite{Kruger:2016:TLB}. \emph{Streaming} is a non-local advection of $f_i$ values along each of the $Q$ discrete directions via a stencil.

In optimized GPU LBM implementations, collision, streaming, and boundary conditions are often fused into a single kernel~\cite{Geier:2017:ETA}. We use the work of Meneghin et al.~\shortcite{Meneghin:2022:NAM} as our baseline since they achieve state-of-the-art results on single- and multi-GPU\@.

\subsection{Improving LBM Scalability}
\label{sec:communications}
\begin{figure}[t]
    \centering
    \subfloat[1D Partition\label{fig:partitionHalo}]{\includegraphics[height=.24\linewidth]{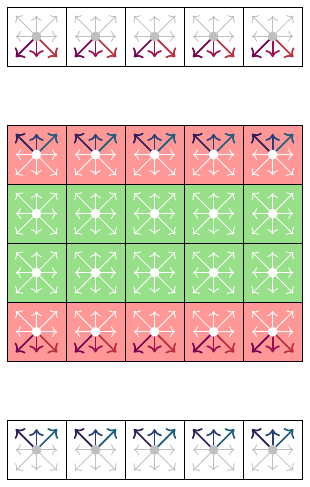}}\hfil
    \subfloat[SoA Layout \label{fig:soaLayout}]{\includegraphics[height=.24\linewidth]{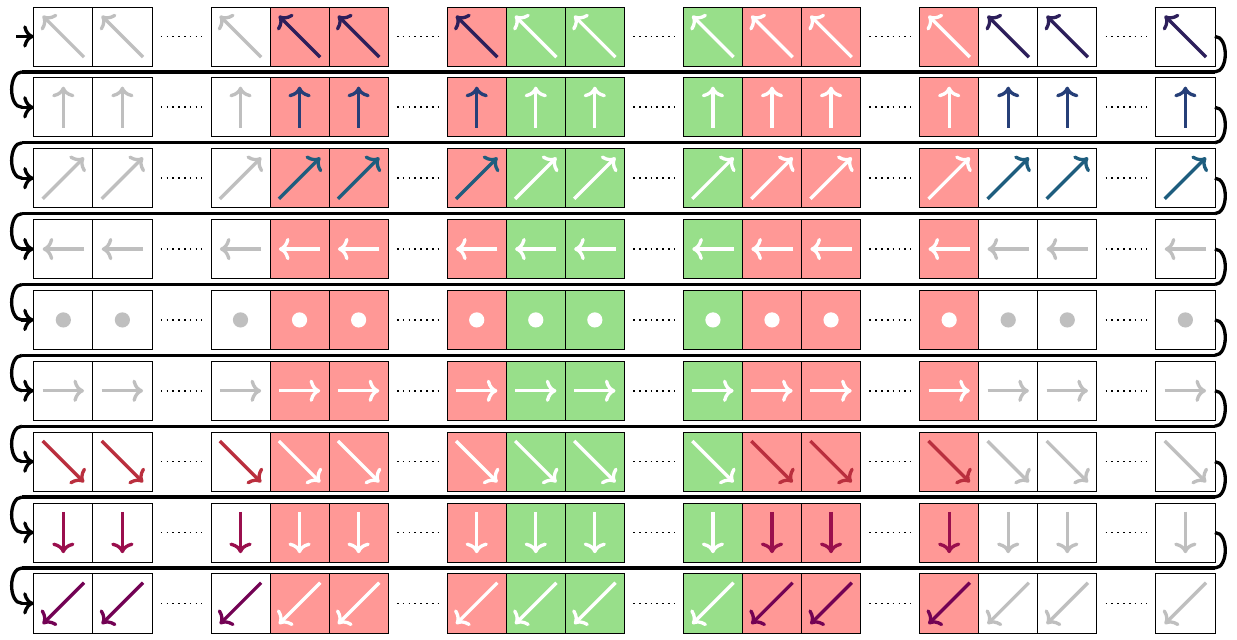}}\hfil\\
    \subfloat[Comp.\ Graph \label{fig:occGraph}] {\includegraphics[height=.21\linewidth]{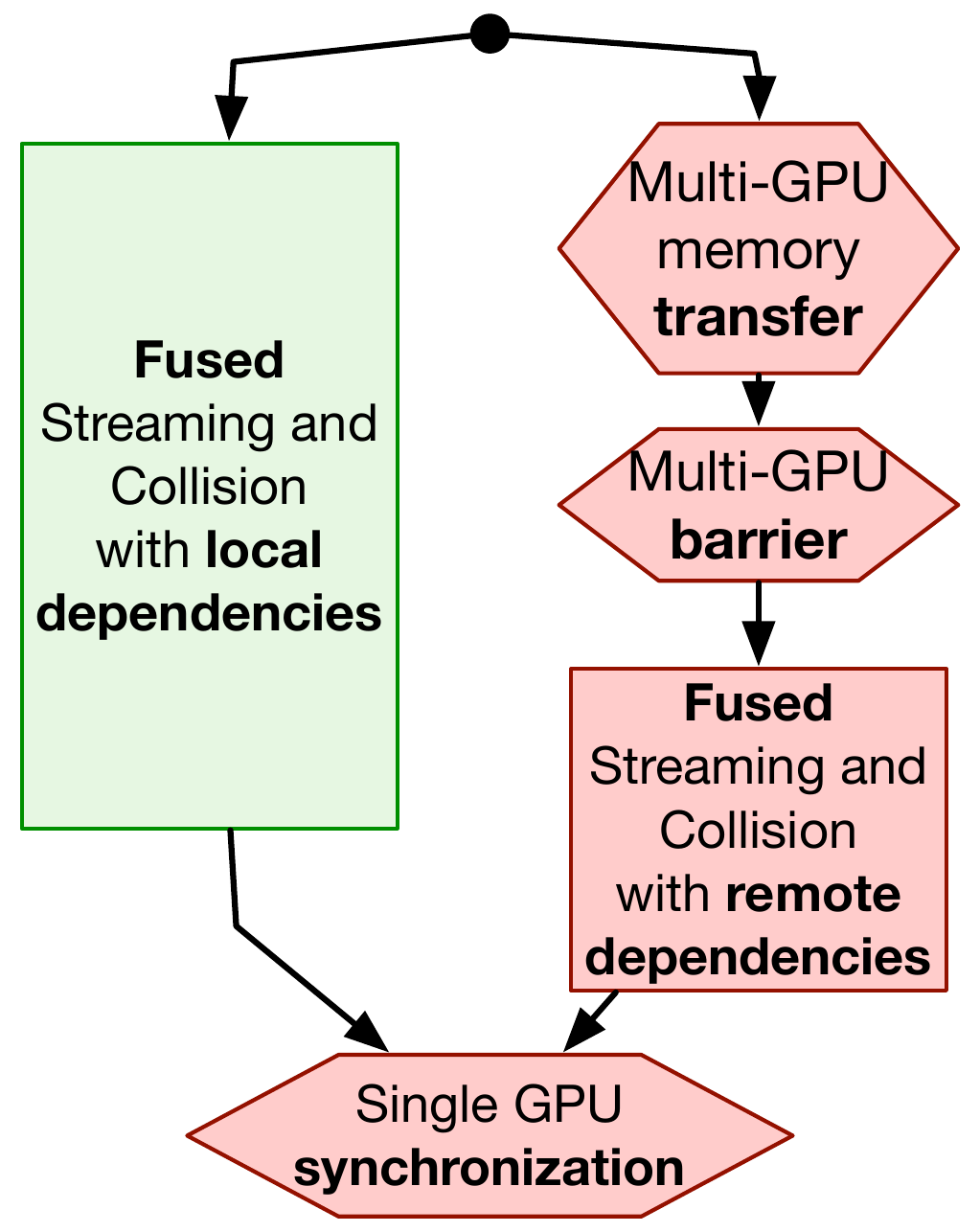}}\hfil
    \subfloat[Disaggregated\label{fig:disComLayout}] {\includegraphics[height=.21\linewidth]{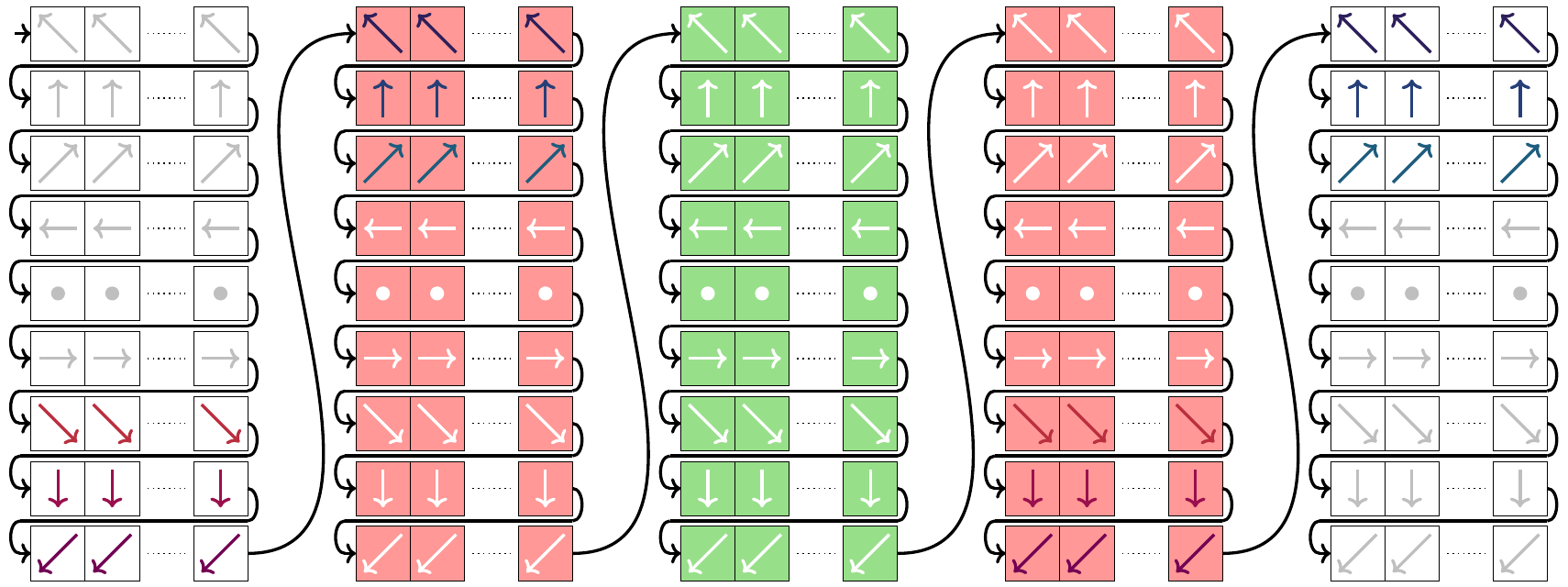}}
    \caption{(a) 1D-partitioned LBM grid: white arrows show local dependencies, and red/blue arrows show dependencies from upper/lower partitions. (c) Computation graph with OCC in the reference implementation. (b,d) SoA and disaggregated SoA layouts for a D2Q9 lattice in 2D, with black arrows indicating memory mapping.
    }
\end{figure}

We evaluate our disaggregated design on a lid-driven cavity flow problem~\cite{Latt:2021:CPM} within a cubic domain, using LBM and a dense voxel representation on single-node multi-GPU systems. We analyze both a theoretical communication model (Section~\ref{sec:dis_sten_pattern}) and runtime performance.

\subsubsection{Reference Implementation}
In single-node multi-GPU systems, inter-GPU communication can occur via PCI or faster interconnects like NVLink. Here, we use native \texttt{cudaMemcpyPeer} for best performance~\cite{Jiri:MGPM:21}. Because these systems typically house up to 8--16 GPUs, we use a 1D partitioning scheme~\cite{Jiri:MGPM:21}, giving each partition at most two neighbors (upper and lower) and enabling efficient zero-copy memory transfers.

Figure~\ref{fig:partitionHalo} shows LBM data dependencies with green \emph{private} voxels (computed locally) and red \emph{shared} voxels (requiring data from an adjacent partition). At the lattice granularity, some populations remain local (white or gray), while others must be exchanged (red or blue). Only certain populations of each shared voxel are transferred, which is a defining feature of the LBM streaming operation.

Efficient OCC is critical for achieving fine-grain scalability in LBM~\cite{Meneghin:2022:NAM}. Figure~\ref{fig:occGraph} show the computation graph where private-voxel computation is overlapped with halo exchanges for shared voxels, hiding latency and improving performance. This OCC-based implementation is our baseline reference.

\subsubsection{Modeling Communication Overhead}
The parameters $\alpha$ and $\beta$ from Eq.~\ref{eq:halo_update_model_1D} vary with LBM lattice and data layout (Table~\ref{tbl:comModel}). Under AoS, all populations in a voxel are contiguous, yielding $\alpha = 2$ transfers (one for each neighbor), but forcing $\beta$ to include unneeded populations. Conversely, in SoA (Figure~\ref{fig:soaLayout}), populations are contiguous \emph{per direction}, increasing $\alpha$ but minimizing $\beta$. Table~\ref{tbl:comModel} shows that AoS has a smaller $\alpha$ but higher $\beta$, while SoA has the opposite trade-off.

\begin{wraptable}[10]{r}{0.5\textwidth}
    \vspace{-2em}
    \centering
    \small
    \begin{tabular}{r|cc|cc|cc}
                        & \multicolumn{2}{c|}{\bf D2Q9} & \multicolumn{2}{c|}{\bf D3Q19} & \multicolumn{2}{c}{\bf D3Q27}                                \\
                        & $\alpha$                      & $\beta$                        & $\alpha$                      & $\beta$ & $\alpha$ & $\beta$ \\
        \midrule \midrule
        {\bf AoS}       & 2                             & 18s                            & 2                             & 38s     & 2        & 54s     \\
        {\bf SoA}       & 6                             & 6s                             & 10                            & 10s     & 18       & 18s     \\
        {\bf Disag SoA} & 2                             & 6s                             & 2                             & 10s     & 2        & 18s     \\
    \end{tabular}
    \caption{LBM communication parameters for Eq.~\ref{eq:halo_update_model_1D}; $s$ is half the shared voxels: $d_x$ in 2D and $d_x \cdot d_y$ in 3D.}
    \label{tbl:comModel}
\end{wraptable}

\subsubsection{Disaggregated Optimization}
We now extend the disaggregated layout introduced for stencil operations on a vector-valued field to fully support \emph{zero-copy} communication by including halo regions. These halos enable direct data sharing without additional staging buffers. Concretely, we define distinct properties to ensure contiguous mappings for each critical region: upper halos, upper boundary voxels, lower boundary voxels, and lower halos. As with our original design, each of these groups is mapped using an SoA layout. Figure~\ref{fig:disComLayout} shows both how the domain is split into groups and how these groups are placed in memory. In this arrangement, any data that needs to be transferred or received resides contiguously---the solid-colored (red or blue) populations in Figure~\ref{fig:disComLayout}. This means each group's data is placed in a continuous block, allowing for a minimal number of bulk transfers. In the D2Q9 example, this disaggregated SoA configuration results in an $\alpha$ value of 2, so each partition sends just one message per neighbor. Meanwhile, $\beta$ remains at 6, representing the exact amount of data required by the stencil. Table~\ref{tbl:comModel} shows that for D3Q19 and D3Q27, the disaggregated layout delivers similarly optimal $\alpha$ and $\beta$ values, consistently outperforming basic AoS or SoA alone. Overall, the disaggregated SoA layout combines AoS-like benefits of minimal transfer operations with SoA's advantage of transferring only the necessary populations. By maintaining zero-copy efficiency, it reduces overhead in inter-partition data exchanges, making it theoretically the most communication-efficient layout for multi-GPU LBM.

\begin{wraptable}[8]{r}{0.55\textwidth}
    \vspace{-2em}
    \centering
    \small
    \begin{tabular}{r|cccc}
        Name     & Arch      & GPUs & Mem  & Interc.  \\
        \midrule
        \midrule
        DGX-A100 & A100-SXM4 & 8    & 40GB & NVLink-2 \\
        AWS p3   & V100-SXM2 & 8    & 16GB & NVLink-1 \\
        AWS g5   & A10       & 8    & 24GB & PCI      \\
    \end{tabular}
    \caption{Machines used in benchmarking.}
    \label{table:commArc}
\end{wraptable}

\subsubsection{Benchmarking}
\begin{figure}[t]
    \centering
    \subfloat[D3Q19, speedup\label{fig:comPerfD38GPUs}]{
        \includegraphics[width=.5\linewidth]{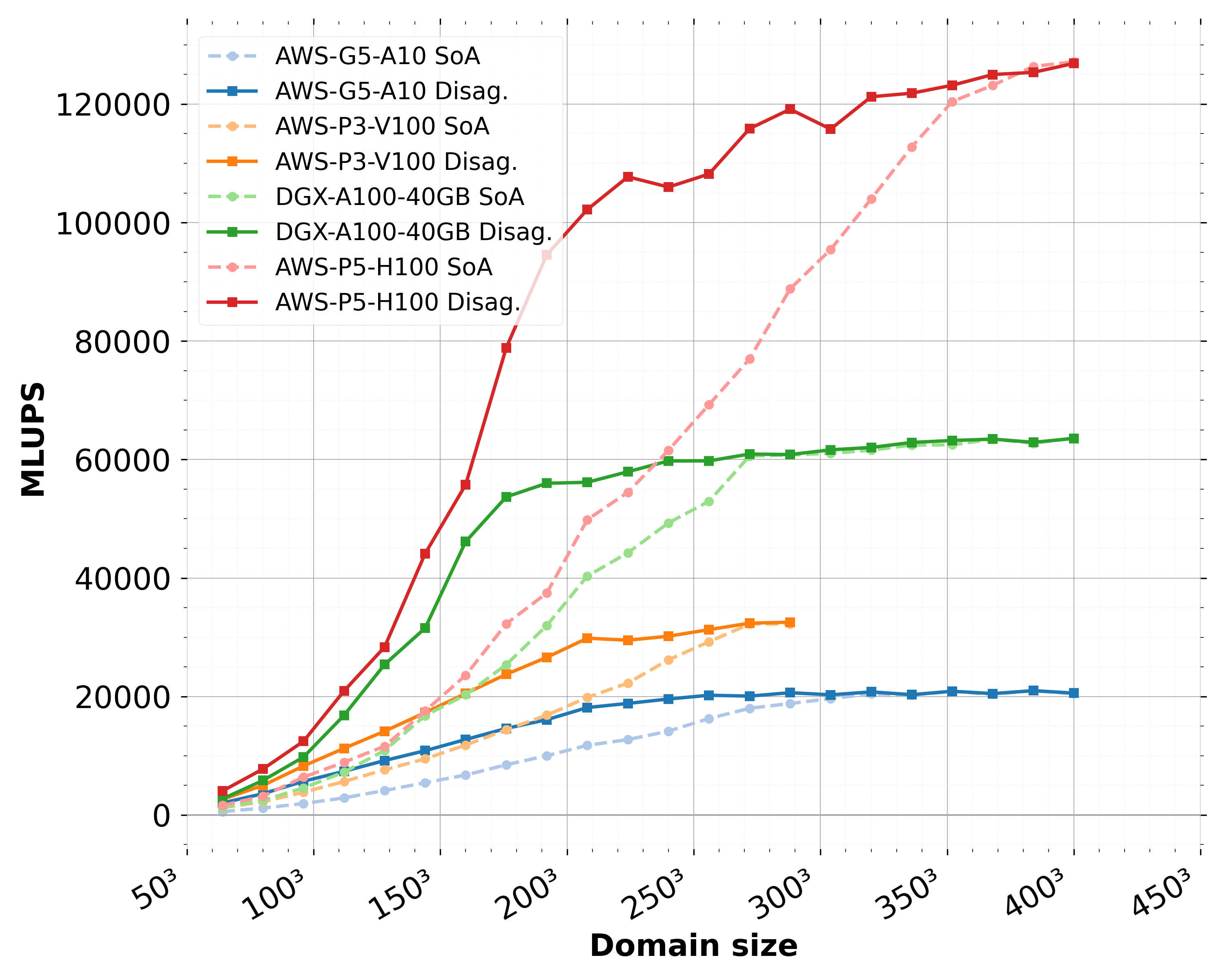}
    }
    \subfloat[D3Q27, strong scaling\label{fig:perf192Strong}]{
        \includegraphics[width=.41\linewidth]{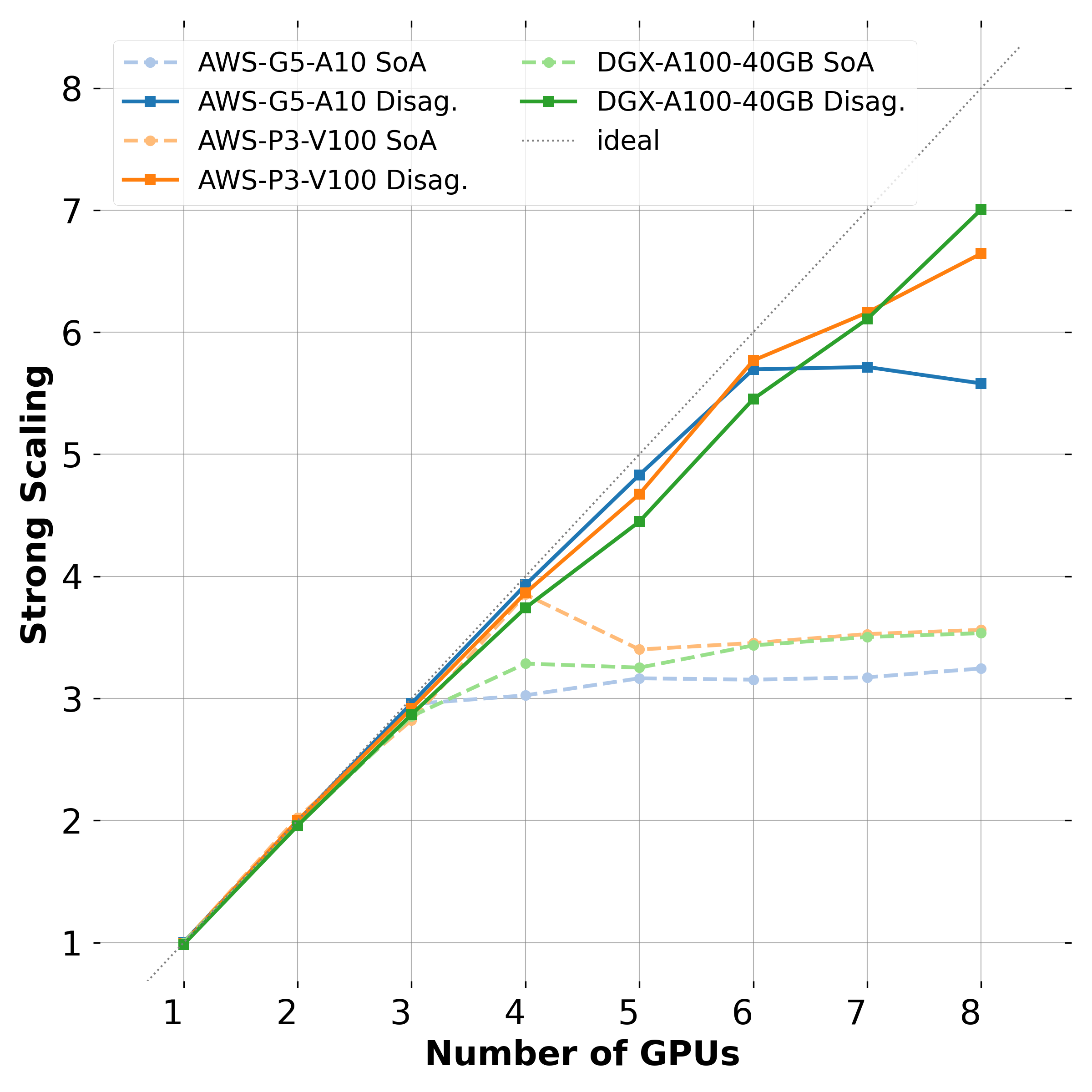}
    }
    \caption{(a) MLUPS performance of disaggregated vs.\ SoA on an 8-GPU lid-driven cavity flow (D3Q19). (b) Strong scaling for D3Q27 on a $192^3$ domain.}
\end{figure}

We measure runtime on a lid-driven cavity flow with a cubic domain, using boundary conditions from Latt et al.~\shortcite{Latt:2021:CPM}. Table~\ref{table:commArc} lists three single-node multi-GPU systems tested, spanning high-end (A100), midrange (A10), and previous-generation (V100) GPUs. A100 and V100 use NVLink; A10 relies on PCI. We exclude AoS due to poor coalesced performance in LBM. Figure~\ref{fig:comPerfD38GPUs} compares disaggregated and SoA layouts for 3D D3Q19 and D3Q27 lattices, using the Million Lattice Updates per Second (MLUPS) metric.

Disaggregated consistently matches or outperforms SoA, especially on smaller domains: up to $4\times$ speedup below $150^3$, about $2.5\times$ from $150^3$--$250^3$, and $1.5\times$ for larger volumes. This reduction in performance improvement for larger domains is well explained by Eq.~\ref{eq:halo_update_model_1D}: with 1D partitioning, the impact of $\beta$ (the amount of data transferred) increases with domain size while the number of private voxels grows cubically with the domain edge length $L$---significantly increasing the amount of computation that can be overlapped with communication. Finally, Figure~\ref{fig:perf192Strong} shows strong scaling for $192^3$ domains. While traditional methods struggle to exceed $3\times$ scaling from a single GPU, the disaggregated layout consistently achieves $6\times$ or more, regardless of GPU architecture.

\subsection{Improving LBM Register Allocation}
\label{sec:registerPressure}

We evaluate the effect of disaggregation on register usage by simulating fluid flow over an obstacle in a cubic domain. The simulation uses a block-sparse grid, with each block containing $4^3$ voxels. This setup resembles a typical wind tunnel, where a \emph{bounce-back} boundary condition~\cite{Paul:BIL:91} is applied to obstacle surfaces. Additionally, an \emph{inflow} boundary condition is applied to one face, and an \emph{outflow} boundary condition is applied to the opposite face.

The bounce-back boundary condition is \emph{register-light}, requiring $2Q$ populations. By contrast, the inflow and outflow faces use a \emph{regularized} boundary condition~\cite{Latt:2008:SVB}, which is \emph{register-heavy}, needing $3Q$ populations. Combining bounce-back and regularized boundaries in one kernel forces resource requirements to accommodate $\max(2Q,3Q)=3Q$, potentially over-allocating registers for most voxels. Because the fraction of regularized voxels is $\mathcal{O}\!\bigl(\tfrac{1}{x}\bigr)$ relative to domain size $x$, this wastes registers on the majority of the grid.

\paragraph{Disaggregated Solution.}
To address this, we classify voxels needing the regularized method as \emph{boundary}, while bounce-back and other voxels are \emph{non-boundary}. Any block containing at least one regularized voxel is marked a \emph{boundary block}; the rest are \emph{non-boundary blocks}. We then assign each category to its own specialized kernel, reducing register pressure for non-boundary blocks. This layout can be implemented by storing boundary blocks contiguously in memory or by using a bitmask to identify them. Both approaches let non-boundary voxels be processed using fewer registers, thus improving occupancy. Table~\ref{table:dissBlock} summarizes these advantages, showing how disaggregation avoids register-related bottlenecks.

\begin{figure}[t]
    \centering
    \begin{minipage}{0.5\textwidth}
        \includegraphics[width=.9\linewidth]{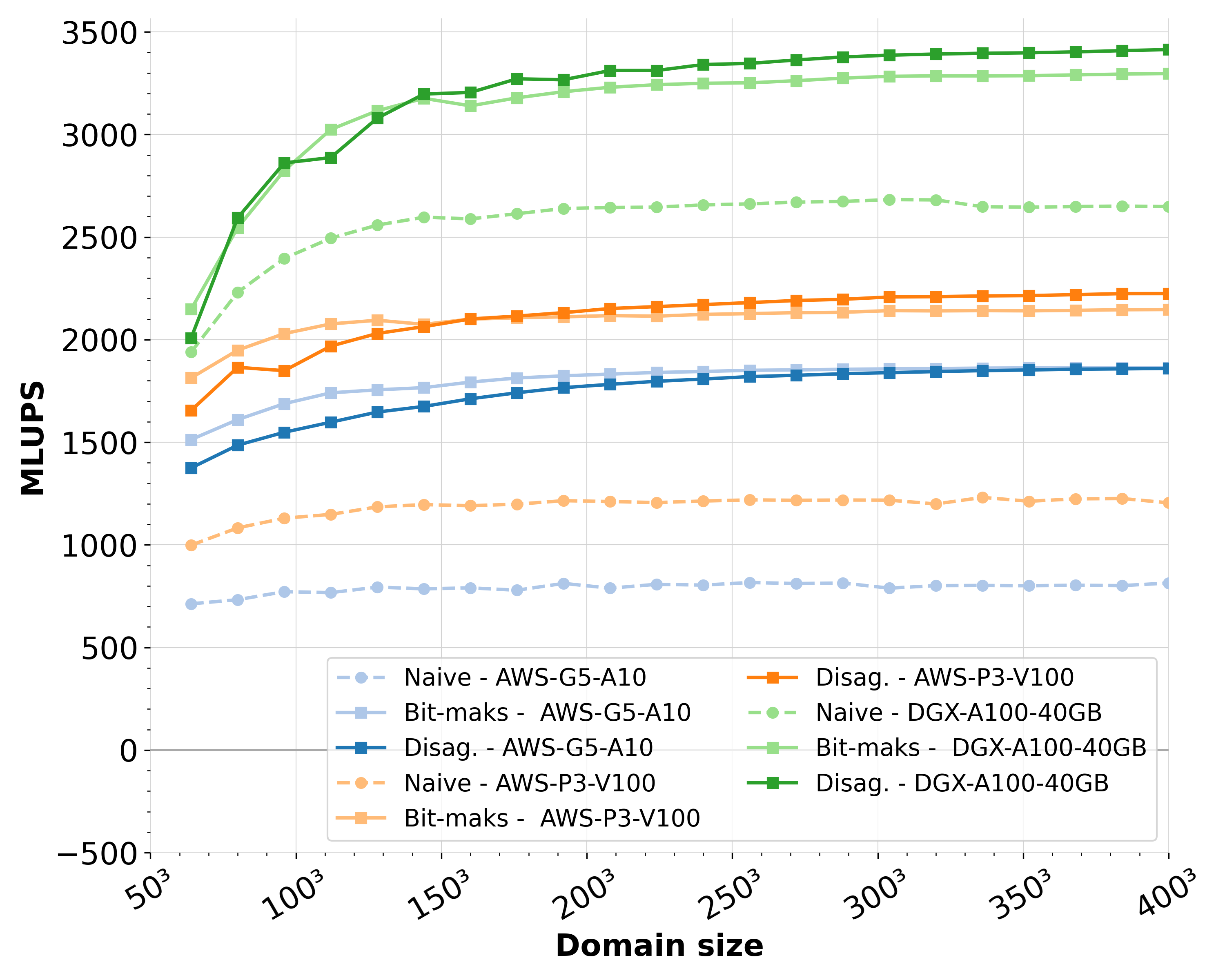}
    \end{minipage}
    \hfill
    \begin{minipage}{0.45\textwidth}
        \caption{\label{fig:plot:regPressure}
        Performance comparison (single GPU) between the naive implementation and the disaggregated approach (both bitmask and continuous block allocations). The domain uses a D3Q27 lattice on a block-sparse grid, with regularized inflow and outflow boundaries~\cite{Latt:2008:SVB}.
        }
    \end{minipage}
\end{figure}

\subsubsection{Benchmarks}
We ran this scenario on a single GPU from each system in Table~\ref{table:commArc}, letting CUDA compiler determine register allocation and spilling. Figure~\ref{fig:plot:regPressure} shows the performance (MLUPS) for D3Q27 across different domain sizes. In every tested case, the disaggregated solutions (bitmask or memory-based) outperform the naive approach. Gains can reach $2\times$ on V100 and A10 and $1.3\times$ on A100. On V100, the naive kernel needs 55 registers—matching the boundary kernel in the disaggregated case—and suffers spills for the entire domain. Under disaggregation, only the smaller boundary block regions invoke the 55-register kernel, while the majority use a lower-resource kernel. For a domain of size 368, the naive approach has $2.2\times$ more L2-DRAM traffic, explaining the $2\times$ speedup.

On A100, all kernels (naive and disaggregated) also require 55 registers. Spilling occurs only in the naive approach, yet the A100's larger L2 cache reduces its overall penalty, limiting speedup to around $1.3\times$. Still, data traffic between L2 and DRAM is $1.3\times$ higher for the naive kernel, matching the measured performance improvement. Finally, regarding memory overhead for boundary conditions, the regularized method stores a $d$-component velocity vector (where $d=2$ in 2D and $d=3$ in 3D). Let $F$ be the floating-point type and $I$ the indexing type. Then $s_w = \text{sizeOf}(F)\cdot d$ and $s_i = \text{sizeOf}(I)$. As noted in Table~\ref{table:dissBlock}, only the disaggregated approach avoids additional storage for boundary voxels; other methods typically allocate boundary-related data throughout the entire domain.

\subsection{Improving Multi-resolution LBM Kernel Fusion}  
\label{sec:kernelfusion}  
\begin{wraptable}[22]{r}{0.55\textwidth}
	\vspace{-2.5em}
	\centering
	\small    
    \begin{tabular}{r|c|r|ccr}
        \toprule
        GPU                  &  Size & Distribution &  Ours    & Baseline  & \textbf{Gain} \\        
        \midrule
        A100& $512^3$ & 77, 4, 0.4        & 6072 & 4824 & \textbf{25\%} \\
        A100& $512^3$ & 73, 3, 0.5, 0.003 & 6018 & 4769 & \textbf{26\%} \\
        
        \midrule
        V100 & $320^3$ & 15, 1, 0.1          & 4421 & 3770 & \textbf{17\%}\\
        V100 & $320^3$ & 15, 1, 0.1, 0.002   & 4422 & 3770 & \textbf{17\%}\\
        V100 & $480^3$ & 53, 4, 0.4          & 5006 & 4047 & \textbf{23\%}\\
        V100 & $480^3$ & 53, 4, 0.3, 0.008   & 5005 & 4050 & \textbf{23\%}\\
        
        \midrule
        A10 & $320^3$ & 15, 1, 0.1, 0.002 & 4083 & 3982 & \textbf{2\%}  \\
        A10 & $480^3$ & 53, 4, 0.3, 0.008 & 4483 & 4306 & \textbf{4\%} \\
        A10 & $512^3$ & 77, 4, 0.4        & 4093 & 3901 & \textbf{4\%} \\
        A10 & $512^3$ & 73, 3, 0.5, 0.003 & 3890 & 3719 & \textbf{4\%} \\
        \bottomrule
    \end{tabular}
    \caption{Performance comparison of our disaggregated multi-resolution LBM approach vs.\ a state-of-the-art baseline~\cite{Mahmoud:OGI:2024} on a lid-driven cavity flow. \emph{Size} is the length of the virtual finest-level box; \emph{Distribution} lists the fraction of active voxels per level (finest to coarsest); \emph{Baseline} is baseline's MLUPS; \emph{Ours} is our MLUPS; and \emph{Gain} is computed as (\textit{Ours}$-$\textit{Baseline})$/$\textit{Baseline}$\times100\%$. \label{table:multiResLbm}}
\end{wraptable}

We evaluated a single-GPU optimized multi-resolution LBM solver \cite{Mahmoud:OGI:2024} enhanced with the disaggregated design. The data structure comprises multiple uniform block-sparse grids (one per resolution level), along with transition metadata to manage inter-level dependencies.

In multi-resolution LBM, each time step entails two key inter-level operations: (1) \emph{Explosion} where collision results at one resolution feed into lower-resolution levels and (2) \emph{Coalescence} where streaming data from higher resolution levels merges into lower-resolution blocks.

These create dependency edges in a graph similar to Figure~\ref{fig:mres:naive}, with operator A for collision and B for streaming. Except at the finest resolution, explosion and coalescence prevent kernel fusion without additional processing.

We classify blocks as: (1) $\mathcal{P}_{\text{uniform}}$ which are blocks that operate uniformly and don't require inter-level synchronization and (2) $\mathcal{P}_{\text{jump}}$ which are blocks containing at least one voxel near a resolution jump, needing explicit explosion/coalescence handling. Using the disaggregated interface, we eliminate unneeded synchronizations at the kernel level. For $\mathcal{P}_{\text{uniform}}$ blocks, we fuse collision and streaming into a single kernel. For $\mathcal{P}_{\text{jump}}$ blocks, the original multi-step execution is preserved to correctly process inter-level data.

Table~\ref{table:multiResLbm} shows a lid-driven cavity flow benchmark at multiple resolutions on three different GPUs. In all configurations, the disaggregated approach outperforms the baseline. On the A100, a high-end architecture, improvements can reach 26\% for domains of size $512^3$. The V100 also achieves up to 23\% on $480^3$ grids. These speedups stem from merging collision and streaming on uniform blocks, thus reducing overhead where inter-level dependencies are not needed.

In contrast, the A10 exhibits smaller gains (2--4\%), largely due to register spilling penalizing performance more acutely on midrange GPUs with smaller cache sizes and lower memory bandwidth. Nonetheless, the disaggregated method consistently demonstrates advantages for larger domains and higher active-voxel counts, confirming its suitability for diverse multi-resolution setups.

\section{Related Work}
\label{sec:background}
While extensive research has been conducted on optimizing stencil computations, most efforts focus on improving data locality. To the best of our knowledge, this work is the first to propose a data structure design methodology aimed at multi-objective optimization, where data locality can be strategically traded off for other performance goals.

\textbf{Optimizing communications via data layout:}
Zhao et al.~\cite{Zhao:ICB:2021} introduced a layout scheme for block representations designed to minimize communication overhead and enable zero-copy communication. Their approach incorporates virtual memory techniques to reduce the impact of indirect indexing which is effective for stencils with a radius that is a multiple of four. However, for stencils with a radius of one, users must resort to time tiling, which can increase message sizes or become infeasible if reductions are involved. While their method demonstrates significant speedups on distributed systems, it does not address the challenges of multi-cardinality fields.

\textbf{Reducing Register Pressure and Spilling:}
Managing register pressure and minimizing spilling are critical challenges in GPU computing. Temporal blocking techniques, e.g., register blocking, serialize one domain dimension to improve data reuse~\cite{Matsumura:AN5D:2020}. Other strategies leverage GPU shared memory to mitigate the impact of spilling~\cite{Sakdhnagool::RegDemotion::2019}. However, no prior work has explored using data structure design to address register pressure.

\textbf{Kernel Fusion:}
Kernel fusion is a well-known optimization for memory-bound problems, as it reduces memory pressure by keeping shared data in registers between consecutive kernels. This technique is widely used in dense LBM implementations~\cite{Latt:2021:CPM} and multi-resolution representations~\cite{Schornbaum:2016:MPA}. However, existing works do not explore leveraging data layouts to facilitate kernel fusion.
\section{Conclusion and Future Work}
\label{sec:conclusion}

Past advances in volumetric computation have helped data-structure designers create layouts that emphasize memory access efficiency. In this work, we introduced \emph{disaggregated design}, a unified framework that broadens the optimization focus beyond data locality to include minimizing multi-GPU data transfers, reducing register pressure, and maximizing kernel fusion. Our analytical models and empirical results confirm the benefits of these objectives, while also clarifying the potential drawbacks of more intricate indexing schemes and slightly compromised locality. Ultimately, the overall gains depend on whether the performance improvements outweigh these costs.

This study represents the first in-depth analysis of disaggregated design and its practical scope. Several directions remain for future exploration: (1) extending the applicability of the disaggregated design to other spatial data structures, e.g., unstructured meshes and hash grids, and (2) exploring additional optimization objectives tailored to diverse computational workloads, e.g., load balancing for particle-based simulations.

\begin{credits}
	\subsubsection{\ackname} The authors express their gratitude to Hesam Salehipour, Mehdi Ataei, and Oliver Hennigh for their invaluable insights into the intricacies of LBM and their support in providing data on H100 architecture. Ahmed Mahmoud acknowledges the generous support of National Science Foundation grant OAC-2403239.
	\subsubsection{\discintname} The authors have no competing interests to declare that are
	relevant to the content of this article.
\end{credits}

\bibliographystyle{splncs04}
\bibliography{gpulbm, comModels, lbm}

\begin{thebibliography}{10}
\providecommand{\url}[1]{\texttt{#1}}
\providecommand{\urlprefix}{URL }
\providecommand{\doi}[1]{https://doi.org/#1}

\bibitem{Asanovic:VPC:2009}
Asanovic, K., Bodik, R., Demmel, J., Keaveny, T., Keutzer, K., Kubiatowicz, J.,
  Morgan, N., Patterson, D., Sen, K., Wawrzynek, J., Wessel, D., Yelick, K.: A
  view of the parallel computing landscape. Communications of the ACM
  \textbf{52}(10),  56--67 (Oct 2009). \doi{10.1145/1562764.1562783}

\bibitem{Bondhugula:2008:APA}
Bondhugula, U., Hartono, A., Ramanujam, J., Sadayappan, P.: A practical
  automatic polyhedral parallelizer and locality optimizer. In: Proceedings of
  the 29th ACM SIGPLAN Conference on Programming Language Design and
  Implementation. pp. 101--113. PLDI '08, Association for Computing Machinery,
  New York, NY, USA (Jun 2008). \doi{10.1145/1375581.1375595}

\bibitem{Culler:1993:LTA}
Culler, D., Karp, R., Patterson, D., Sahay, A., Schauser, K.E., Santos, E.,
  Subramonian, R., von Eicken, T.: Log{P}: towards a realistic model of
  parallel computation. SIGPLAN Not.  \textbf{28}(7),  1--12 (Jul 1993).
  \doi{10.1145/173284.155333}

\bibitem{Endo:2018:ART}
Endo, T.: Applying recursive temporal blocking for stencil computations to
  deeper memory hierarchy. In: 2018 IEEE 7th Non-Volatile Memory Systems and
  Applications Symposium (NVMSA). pp. 19--24 (Nov 2018).
  \doi{10.1109/NVMSA.2018.00016}

\bibitem{Frigo:2012:COA}
Frigo, M., Leiserson, C.E., Prokop, H., Ramachandran, S.: Cache-oblivious
  algorithms. ACM Trans. Algorithms  \textbf{8}(1) (Jan 2012).
  \doi{10.1145/2071379.2071383}

\bibitem{Geier:2017:ETA}
Geier, M., Sch{\"o}nherr, M.: {E}soteric {T}wist: An efficient in-place
  streaming algorithmus for the lattice boltzmann method on massively parallel
  hardware. Computation  \textbf{5}(2), ~19 (Mar 2017).
  \doi{10.3390/computation5020019}

\bibitem{Jiri:MGPM:21}
Kraus, J.: Multi-{GPU} programming models,
  \url{https://www.nvidia.com/en-us/on-demand/session/gtcfall21-a31140/}

\bibitem{Kruger:2016:TLB}
Kr{\"u}ger, T., Kusumaatmaja, H., Kuzmin, A., Shardt, O., Silva, G., Viggen,
  E.M.: The {L}attice {B}oltzmann method. Springer Cham, 1st edn. (2016).
  \doi{10.1007/978-3-319-44649-3}

\bibitem{Lagrava:2012:AIM}
Lagrava, D., Malaspinas, O., Latt, J., Chopard, B.: Advances in multi-domain
  lattice {B}oltzmann grid refinement. Journal of Computational Physics
  \textbf{231},  4808--4822 (may 2012). \doi{10.1016/j.jcp.2012.03.015}

\bibitem{Latt:2008:regularized}
Latt, J., Chopard, B., Malaspinas, O., Deville, M., Michler, A.: Straight
  velocity boundaries in the lattice {B}oltzmann method. Physical Review E
  \textbf{77}(5),  056703 (2008)

\bibitem{Latt:2008:SVB}
Latt, J., Chopard, B., Malaspinas, O., Deville, M., Michler, A.: Straight
  velocity boundaries in the lattice {B}oltzmann method. Physical Review E
  \textbf{77} (May 2008). \doi{10.1103/PhysRevE.77.056703}

\bibitem{Latt:2021:CPM}
Latt, J., Coreixas, C., Beny, J.: Cross-platform programming model for
  many-core lattice {B}oltzmann simulations. PLOS ONE  \textbf{16}(4),  1--29
  (04 2021). \doi{10.1371/journal.pone.0250306}

\bibitem{Paul:BIL:91}
Lavall\'{e}e, P., Boon, J.P., Noullez, A.: Boundaries in lattice gas flows.
  Physica {D}: Nonlinear Phenomena  \textbf{47}(1),  233--240 (1991).
  \doi{10.1016/0167-2789(91)90294-J}

\bibitem{Mahmoud:OGI:2024}
Mahmoud, A.H., Salehipour, H., Meneghin, M.: Optimized {GPU} implementation of
  grid refinement in lattice {B}oltzmann method. In: Proceedings of the 38th
  IEEE International Parallel and Distributed Processing Symposium. pp.
  398--407 (Jul 2024). \doi{10.1109/IPDPS57955.2024.00042}

\bibitem{Matsumura:AN5D:2020}
Matsumura, K., Zohouri, H.R., Wahib, M., Endo, T., Matsuoka, S.: {AN5D}:
  automated stencil framework for high-degree temporal blocking on {GPU}s. In:
  Proceedings of the 18th ACM/IEEE International Symposium on Code Generation
  and Optimization. pp. 199--211. CGO '20, Association for Computing Machinery,
  New York, NY, USA (2020). \doi{10.1145/3368826.3377904}

\bibitem{Meneghin:2022:NAM}
Meneghin, M., Mahmoud, A.H., Jayaraman, P.K., Morris, N.J.W.: {N}eon: A
  multi-{GPU} programming model for grid-based computations. In: Proceedings of
  the 36th IEEE International Parallel and Distributed Processing Symposium.
  pp. 817--827 (Jun 2022). \doi{10.1109/IPDPS53621.2022.00084}

\bibitem{Sakdhnagool::RegDemotion::2019}
Sakdhnagool, P., Sabne, A., Eigenmann, R.: Optimizing {GPU} programs by
  register demotion: poster. In: Proceedings of the 24th Symposium on
  Principles and Practice of Parallel Programming. pp. 405--406. PPoPP '19,
  Association for Computing Machinery, New York, NY, USA (2019).
  \doi{10.1145/3293883.3297859}

\bibitem{Schornbaum:2016:MPA}
Schornbaum, F., R\"{u}de, U.: Massively parallel algorithms for the lattice
  boltzmann method on nonuniform grids. SIAM Journal on Scientific Computing
  \textbf{38},  C96--C126 (2016). \doi{10.1137/15M1035240}

\bibitem{Tran:2017:POO}
Tran, N.P., Lee, M., Hong, S.: Performance optimization of 3d lattice
  {B}oltzmann flow solver on a {GPU}. Scientific Programming  \textbf{2017}(1)
  (Jan 2017). \doi{10.1155/2017/1205892}

\bibitem{Wang:2020:AMP}
Wang, X., Qiu, Y., Slattery, S.R., Fang, Y., Li, M., Zhu, S.C., Zhu, Y., Tang,
  M., Manocha, D., Jiang, C.: A massively parallel and scalable multi-{GPU}
  material point method. ACM Trans. Graph.  \textbf{39}(4) (Aug 2020).
  \doi{10.1145/3386569.3392442}

\bibitem{Wittmann:2013:COD}
Wittmann, M., Zeiser, T., Hager, G., Wellein, G.: Comparison of different
  propagation steps for lattice {B}oltzmann methods. Comput. Math. Appl.
  \textbf{65}(6),  924--935 (Mar 2013). \doi{10.1016/j.camwa.2012.05.002}

\bibitem{Wonnacott:2000:UTS}
Wonnacott, D.: Using time skewing to eliminate idle time due to memory
  bandwidth and network limitations. In: Proceedings 14th International
  Parallel and Distributed Processing Symposium. pp. 171--180. IPDPS 2000 (May
  2000). \doi{10.1109/IPDPS.2000.845979}

\bibitem{Wulf:HTM:1995}
Wulf, W.A., McKee, S.A.: Hitting the memory wall: implications of the obvious.
  ACM SIGARCH Computer Architecture News  \textbf{23}(1),  20--24 (Mar 1995).
  \doi{10.1145/216585.216588}

\bibitem{Zhao:ICB:2021}
Zhao, T., Hall, M., Johansen, H., Williams, S.: Improving communication by
  optimizing on-node data movement with data layout. In: Proceedings of the
  26th ACM SIGPLAN Symposium on Principles and Practice of Parallel
  Programming. pp. 304--317. PPoPP '21, Association for Computing Machinery,
  New York, NY, USA (2021). \doi{10.1145/3437801.3441598}

\end{thebibliography}

\end{document}